\documentclass[aps,showpacs,amsmath,amssymb,pre,superscriptaddress,twocolumn]{revtex4}
\usepackage{graphicx}
\usepackage{epsfig}
\usepackage{ulem}
\usepackage{amsmath}
\usepackage{amssymb}
\usepackage{color}

\begin{document}

\title{Theory of ferromagnetism driven by superexchange in dilute magnetic semiconductors}

\author{C. Simserides}
\email{csimseri@phys.uoa.gr}
\affiliation{Physics Department, University of Athens, GR-15784 Athens, Greece}
\author{J.A. Majewski}
\affiliation{Institute of Theoretical Physics, Faculty of Physics, University of Warsaw, PL-00 681 Warszawa, Poland}
\author{K.N. Trohidou}
\affiliation{Institute for Advanced Materials, Physicochemical Processes, Nanotechnology and Microsystems, NCSR Demokritos, GR-15310 Athens, Greece}
\author{T. Dietl}
\affiliation{Institute of Theoretical Physics, Faculty of Physics, University of Warsaw, PL-00 681 Warszawa, Poland}
\affiliation{Institute of Physics, Polish Academy of Sciences, PL-02 668 Warszawa, Poland}
\affiliation{WPI-Advanced Institute for Materials Research, Tohoku University, Sendai 980-8577, Japan}

\date{\today}% It is always \today, today, but any date may be explicitly specified

\begin{abstract}
Magnetic properties of Ga$_{1-x}$Mn$_x$N are studied theoretically by employing a tight binding approach to determine exchange integrals $J_{ij}$ characterizing the coupling between Mn spin pairs located at distances $R_{ij}$ up to the 16th cation coordination sphere in zinc-blende GaN. It is shown that for a set of experimentally determined input parameters there are no itinerant carriers and the coupling between localized Mn$^{3+}$ spins in GaN proceeds via superexchange that is ferromagnetic for all explored $R_{ij}$ values. Extensive Monte Carlo simulations serve to evaluate the magnitudes of Curie temperature $T_\mathrm{C}$ by the cumulant crossing method. The theoretical values of $T_\mathrm{C}(x)$ are in quantitative agreement with the experimental data that are available for Ga$_{1-x}$Mn$_x$N with randomly distributed Mn$^{3+}$ ions with the concentrations $0.01 \leq x \leq 0.1$.
\end{abstract}

%75.50.Pp 	Magnetic semiconductors 
%75.30.Et 	Exchange and superexchange interactions (see also 71.70.Gm Exchange interactions)
%75.40.Cx 	Static properties (order parameter, static susceptibility, heat capacities, critical exponents, etc.) 
%61.72.uj 	III-V and II-VI semiconductors

\maketitle

%%%%%%%%%%%%%%%%%%%%%%%%%%%%%%%
\section{Introduction}%%%%%%%%%
\label{intro}%%%%%%%%%%%%%%%%%%
%%%%%%%%%%%%%%%%%%%%%%%%%%%%%%%
Dilute magnetic insulators constitute an emerging class of magnetic semiconductors in which rather than the $p\texttt{-}d$ Zener mechanism~\cite{Dietl:2013_arXiv}, {\it ferromagnetic} superexchange accounts for the coupling between diluted transition metal (TM) spins~\cite{bonanni:2011}. Due to its compatibility with III-nitrides that have reached the status of the most important semiconductors next to Si, particularly attractive is Ga$_{1-x}$Mn$_x$N, in which there are no itinerant holes but nevertheless ferromagnetic spin-spin interactions are observed~\cite{bonanni:2011}.

Recent progress in epitaxy, contamination-free processing, and (nano)characterization~\cite{bonanni:2011,Stefanowicz:2010_PRB,sawicki:2012,kunert:2012} allowed the preparation of Ga$_{1-x}$Mn$_x$N films with the randomly distributed Mn$^{3+}$ ions up to $x = 0.1$ showing  $T_\mathrm{C}$ up to about 13~K~\cite{stefanowicz:2013} despite the absence of itinerant carriers. A high degree of crystallinity, a random distribution of the Mn ions, and a weak degree of compensation by residual donors were checked in these samples by a range of electron microscopy, synchrotron radiation, ion beam, optical, and magnetic resonance techniques~\cite{bonanni:2011,Stefanowicz:2010_PRB,kunert:2012}.

The above experimental results were successfully described by the present authors within a tight binding approach (TBA)  and Monte Carlo (MC) simulations~\cite{sawicki:2012,stefanowicz:2013} indicating that {\it ferromagnetic} superexchange accounts for ferromagnetism in Ga$_{1-x}$Mn$_x$N with randomly distributed localized Mn$^{3+}$ spins. Within this model, a change of the Mn charge from 3+ to 2+ results in antiferromagnetic superexchange that dominates in intrinsic II-VI Diluted Magnetic Semiconductors (DMSs) such as Cd$_{1-x}$Mn$_x$Te~\cite{spin-glass-freezing,lipinska:2009} and, presumably, in Ga$_{1-x}$Mn$_x$N containing a sizable concentration of compensating donor defects. Here we discuss briefly the theoretical approach allowing to understand ferromagnetism in Ga$_{1-x}$Mn$_x$N.

%%%%%%%%%%%%%%%%%%%%%%%%%%%%
\section{Theory}%%%%%%%%%%%%
\label{theory}%%%%%%%%%%%%%%
%%%%%%%%%%%%%%%%%%%%%%%%%%%%
Blinowski, Kacman and Majewski~\cite{blinowski:1996} developed a theory of superexchange interactions between substitutional TM ions in zinc-blende semiconductor compounds. The magnetic ions were described in terms
of Parmenter's generalization of the Anderson Hamiltonian for the relevant electronic configuration of the TM, taking into
account the Jahn-Teller distortion,
whereas the host band structure was modeled by the $sp^3s^*$ TBA. They calculated numerically the energy of exchange interaction between Cr$^{2+}$ ions for zinc chalcogenides (ZnS, ZnSe, ZnTe) and found that the superexchange is {\it ferromagnetic}.

We realized that this theory can be adopted for Ga$_{1-x}$Mn$_x$N \cite{sawicki:2012}, where holes introduced by the Mn acceptors are tightly bound, so that no insulator-to-metal transition occurs up to at least $x = 0.1$~\cite{stefanowicz:2013}. Under these conditions, there are no itinerant carriers and the Mn ions assume the $3+$ charge state characterized by the electron configuration identical to Cr$^{2+}$ substitutional cations in Cr-doped zinc-blende II-VI zinc chalcogenides~\cite{blinowski:1996}.
In addition,  the lack of mixed valence (all Mn ions are in the same 3+ charge state), precludes the presence of double exchange.
In this situation, the superexchange accounts for the spin-spin interactions. Its sign is determined by the Anderson-Goodenough-Kanamori rules.

GaN has a wurtzite (wz) structure with $a =$ 0.3188 nm and $c =$ 0.5185 nm.
Hence, to obtain the zinc-blende (zb) analogue with identical density of cation sites, we take the lattice parameter, $a_0 = (\sqrt{3} a^2 c)^{1/3} = $ 0.45 nm. Here $c/a \approx 1.626$,
whereas for the "perfect" wz structure $c/a = (\frac{8}{3})^{1/2} \approx 1.633$.
In the equivalent zb structure, in the fcc cation sublattice, we name $D$ "the diameter of hard touching spheres", i.e. $D = \frac{\sqrt{2}}{2} a_0$. In the fcc lattice, the distance of the $n$th nearest neighbors (NNs) up to the 16th ones  \cite{zallen:1983}, $r_n = D \sqrt{n}, n \leq 13$ and $r_n = D \sqrt{n+1}, 14 \leq n \leq 16$.
%\begin{equation}
%r_n = \left\{
%\begin{array}{ll}
%D \sqrt{n},   & \quad n \leq 13 \\
%D \sqrt{n+1}, & \quad 14 \leq n \leq 16 \end{array} \right..
%\end{equation}
There are no neighbors at distance $D \sqrt{14}$ and further away at distance $D \sqrt{30}$. We mention that fcc and hcp (corresponding to the wz structure) have the same number of 1st and 2nd nearest neighbors, i.e. 12 and 6, respectively, as well as at the same distances, i.e. $D$ and $D\sqrt{2}$, respectively. The differences start at greater distances, e.g. fcc has 24 3rd nearest neighbors at distance $D\sqrt{3}$, while hcp has 2 3rd nearest neighbors at a distance $D \sqrt{\frac{8}{3}}$~\cite{pollack:1964}. However, overall, fcc and hcp have the same atomic packing factor $\pi/\sqrt{18} \approx 0.74$. Moreover, hcp and fcc have almost identical {\it bond} percolation threshold ($p_c^b$) and {\it site} percolation threshold ($p_c^s$) for nearest neighbors~\cite{lorenz:2000}. Hence, wz-GaN can be approximated by zb-GaN in a first decent approach.

%%%%%%%%%%%%%%%%%%%%%%%%%%%%%%%%%%%%%%%%%
\subsection{Tight binding approach}%%%%%%
\label{TB}%%%%%%%%%%%%%%%%%%%%%%%%%%%%%%%
%%%%%%%%%%%%%%%%%%%%%%%%%%%%%%%%%%%%%%%%%
The calculations of the exchange energies $J_{ij}$ are performed for zinc-blende GaN, i.e. the cationic sublattice is fcc. The host band structure is modeled by the $sp^3s^*$ TBA, employing the established parametrization for GaN in the
cubic approximation \cite{ferhat:1996}. The integrals over the Brillouin zone are performed using 2048 $k$-points. This guarantees that $J_{ij}$ are computed with an accuracy of 0.0002 K.

The magnetic ions are described in terms
of the Parmenter Hamiltonian taking into
account the Jahn-Teller distortion \cite{gosk:2005,StroppaKresse:2009}. Since the effect of spin-orbit splitting is small in the valence band of GaN, the spin-dependent interaction between two
Ga-substitutional Mn spins is described by a scalar Heisenberg coupling
\begin{equation}
H_{ij}^{\gamma\delta} = - k_{B} J_{ij}^{\gamma\delta} \textbf{S}_i \cdot \textbf{S}_j.\vspace{-0.2cm}
\end{equation}
$\gamma$ and $\delta$ denote the  $t_{2}$ orbital ($xy$, $xz$, or $yz$) which is empty at the Mn$^{3+}$ ions $i$ and $j$, respectively. $k_{B}$ is the Boltzmann constant. The magnitudes of $J_{ij}^{\gamma\delta}$ are evaluated within the fourth-order perturbation theory in $V_{pd}$ for all possible
orbital configurations $\gamma$ and $\delta$. Similarly to the case of Cr$^{2+}$ ions in II-VI compounds \cite{blinowski:1996} the main contribution originates from quantum hopping involving occupied $t_{2}$ orbitals at the one Mn$^{3+}$ ion and the empty orbital at the other Mn$^{3+}$ ion.
For the orbital configurations in question, we find that the averaged over the shell interaction is ferromagnetic at all distances.

The parameters of the model \cite{blinowski:1996} are taken from results of optical \cite{graf:2003,han:2005}
as well as photoemission and soft x-ray absorption spectroscopies
of Ga$_{1-x}$Mn$_x$N \cite{hwang:2005}.
The charge transfer energy between the Mn ion and the top of the valence band, Mn$^{2+}$ $\rightarrow$ Mn$^{3+}$, $e_1 = - $ 1.8 eV \cite{graf:2003,han:2005}, which together with the on-site
correlation energy for Mn$^{3+}$ ions \cite{graf:2003,han:2005} $U =$ 1 eV, and the on-site exchange energy for Mn${^2+}$ ions, $ \Delta = E(S = 5/2) - E(S = 3/2) = $ 2 eV, leads to $e_2 = $ 4.8 eV, where the uncertainty on
the relevant energies $e_1$ and $e_2$ is, presumably, of the order of
$\pm$ 0.5 eV. The magnitude of the $p$-$d$ hybridization energy is
$V_{pd\sigma} = - 1.5 \pm 0.1$~eV \cite{hwang:2005}. Additionally, we take \cite{ferhat:1996}
$V_{sp\sigma} =  -$1.5 eV,
$V_{pp \pi} = $ 0.675 eV, and
$V_{pp \sigma} = -$1.62 eV.

\begin{figure}[t!]
\centering
\includegraphics[width=8cm]{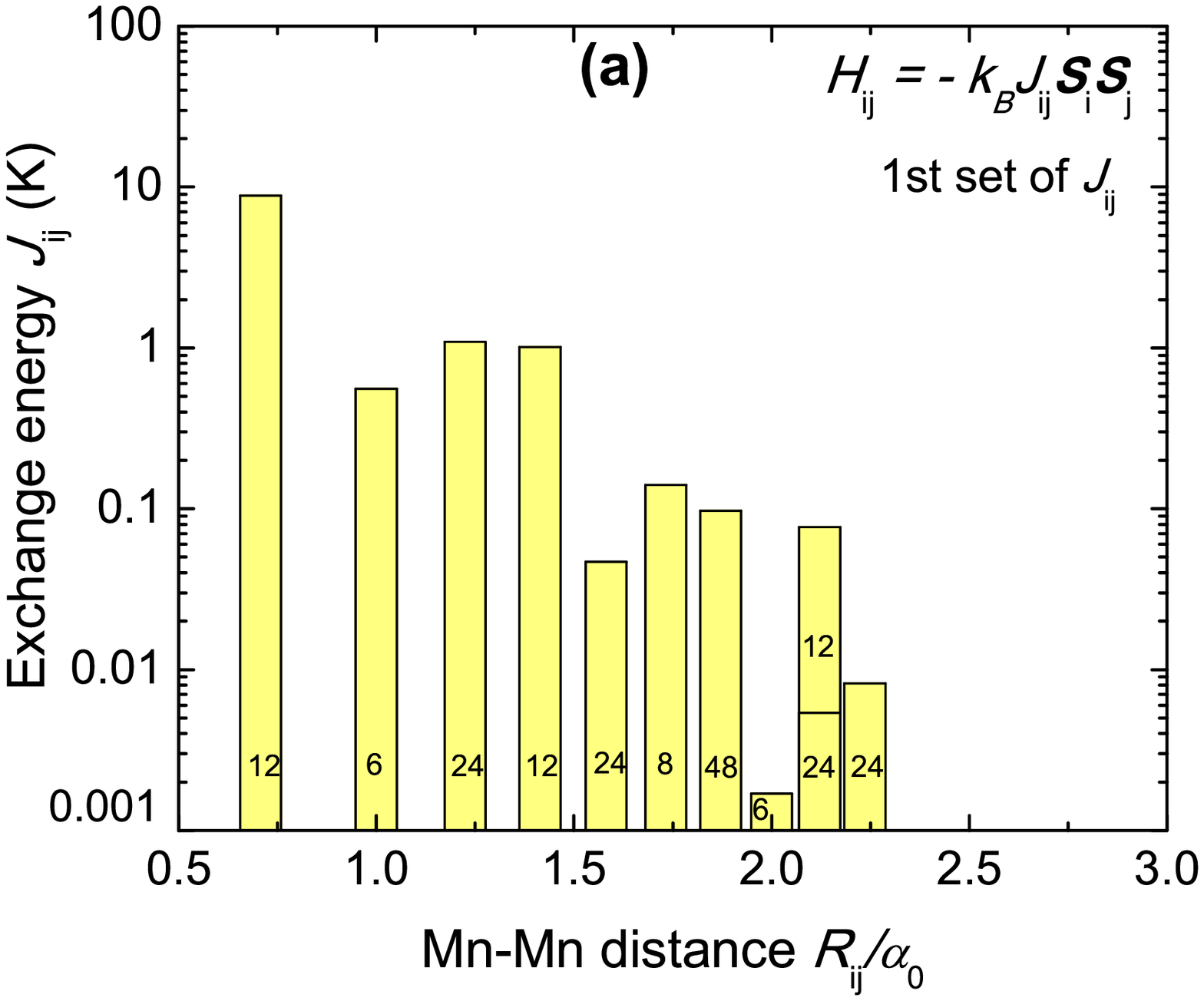}
\includegraphics[width=8cm]{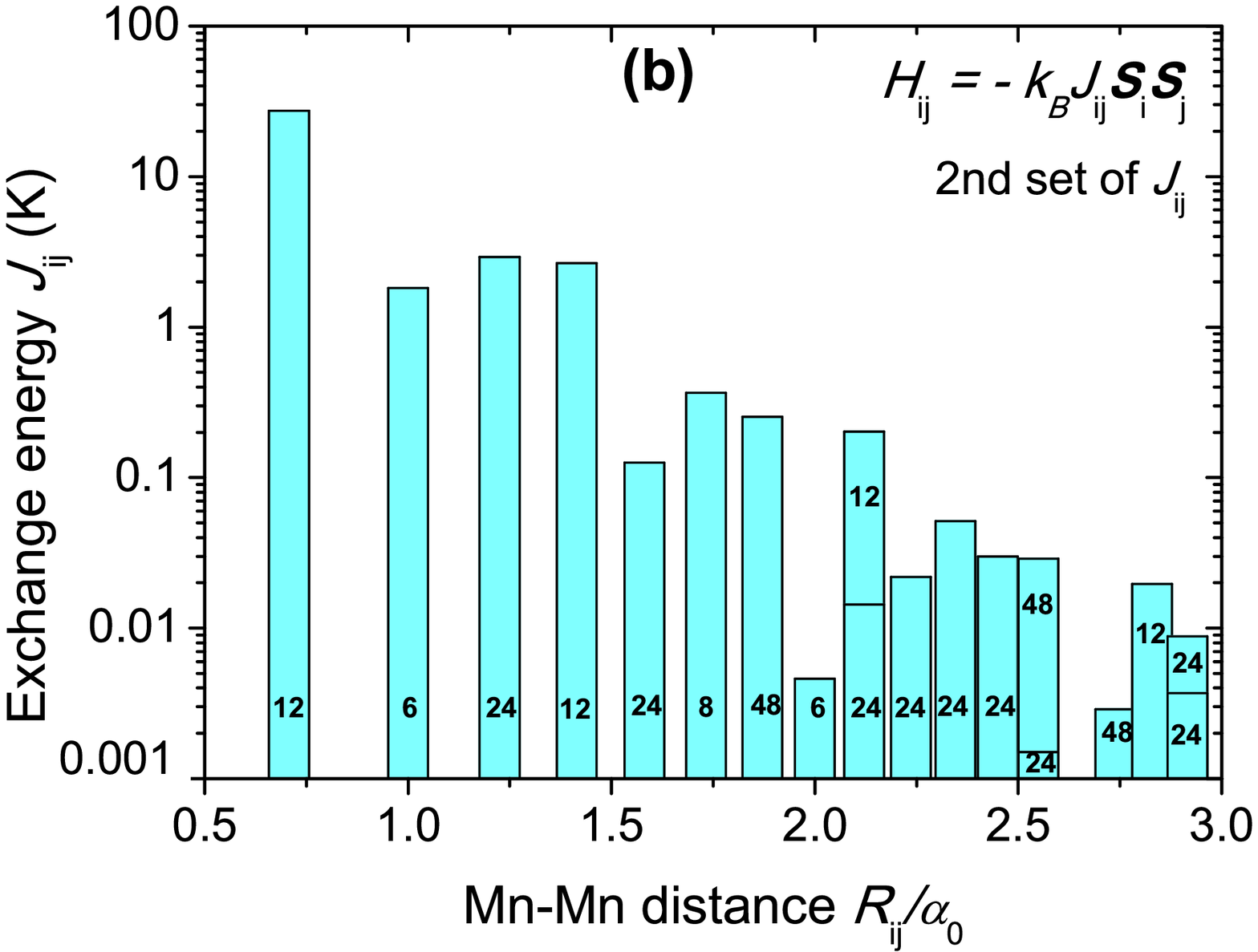}
\caption{ Tight binding exchange energies $J_{ij}$ for zinc-blende GaN vs. Mn-Mn distances $R_{ij}$ in the units of the lattice parameter $a_0$. The numbers of cation sites corresponding to particular shells are also shown. The 1st set of $J_{ij}$ values shown in (a) was used in Ref.~\cite{sawicki:2012}. The 2nd set of $J_{ij}$ shown in (b) was used in Ref.~\cite{stefanowicz:2013}.}
\label{Fig:JsTC}
\end{figure}

To compare quantitatively the theoretical and experimental results, we assume a statistical distribution of
directions corresponding to tetragonal Jahn-Teller distortions and determine the average value of the exchange energy $J_{ij}$ characterizing the coupling of Mn$^{3+}$ pairs at a given distance $R_{ij}$ in the fcc cation sublattice. This way we obtain the values of $J_{ij}$ shown in Fig.~\ref{Fig:JsTC} together with the number of cations corresponding to particular coordination shells. The 1st set of exchange energies corresponds to 10 NNs (Fig.~\ref{Fig:JsTC}(a) \cite{sawicki:2012}). The 2nd set (Fig.~\ref{Fig:JsTC}(b)) is obtained \cite{stefanowicz:2013} for the 16th NNs and reducing the magnitude of $e_2$ from 4.8 to 4.4 eV, i.e. within its expected experimental uncertainty. We find a better agreement between theoretical and experimental $T_\mathrm{C}(x)$ values
with the 2nd set of exchange energies (see, Sec.~\ref{RD}).

%%%%%%%%%%%%%%%%%%%%%%%%%%%%%%%%%%%%%%%%%
\subsection{Monte Carlo simulations}%%%%%
\label{MC}%%%%%%%%%%%%%%%%%%%%%%%%%%%%%%%
%%%%%%%%%%%%%%%%%%%%%%%%%%%%%%%%%%%%%%%%%
We use the Heisenberg Hamiltonian
\begin{equation}\label{hdd}
\mathcal{H}_{dd} = - k_{B} \sum_{i<j}  J_{ij} \textbf{S}_{i} \cdot \textbf{S}_{j}.
\end{equation}
We treat the Mn$^{+3}$ spins $\textbf{S}_i$, $\textbf{S}_j$ as classical vectors with norm $S =$ 2.
Mean values (i.e. per Mn ion) are denoted by $\overline{\cdots}$, statistical averages~\cite{Newman+Barkema:1999} by $\langle \cdots \rangle$.
At each MC sweep, we calculate the mean spin projections ($l=x,y,z,$) and the mean spin norm
%\begin{equation}\label{S-per-particle-components+norm}
$\overline{S_l} = \sum_{i=1}^N S_{il}/N$ and
$\overline{S} = \sum_{i=1}^N S_{i}/N$.
%\end{equation}
$N = N_{\textrm{Mn}}$ is the number of Mn ions.
%\begin{equation}\label{S-st-av-components+norm}
$\langle S_l \rangle = \sum_{n=1}^{n_t}\overline{S_l}/n_t$ and
$\langle S   \rangle = \sum_{n=1}^{n_t}\overline{S}/n_t$.
%\end{equation}
$n$ ($n_t$) denotes successive (the total number of) MC sweeps used for the statistical average.
Similarly, for any integer $p$ one could define
%\begin{equation}\label{S-st-av-normatp}
$\langle S^p \rangle = \sum_{n=1}^{n_t}\overline{S}^p/n_t$.
%\end{equation}
The spin susceptibility components per spin ($\chi_{S_l}$) and the spin susceptibility per spin ($\chi_{S}$)~\cite{Newman+Barkema:1999} are:
%\begin{equation}\label{suscept-components+norm}
$\chi_{S_l} = (N/T) [\langle S_l^2 \rangle - \langle S_l \rangle^2] $ and
$\chi_{S  } = (N/T) [\langle S^2   \rangle - \langle S   \rangle^2] $.
%\end{equation}
We use the Metropolis algorithm. In one MC sweep all Mn spins are rotated.
We usually keep 2000 initial MC sweeps to thermalize the system.
The typical total number of MC sweeps is 120000.
Our simulation cubes ($L \times L \times L$) have typically
$L = 40 a_0 = $ 18 nm,
$L = 50 a_0 = $ 22.5 nm,
$L = 60 a_0 = $ 27 nm.
We use the Mersenne Twister (pseudo)random number generator due to its huge period and its very high order of dimensional equidistribution \cite{Matsumoto:1998}.
Usually $T_{\mathrm{C}}$ is identified with the the peak of the susceptibility $\chi_{S}$.
However, a subtle point is that in reality the results of any MC simulation depend on the size of the system, especially, for small system sizes. One avenue out of this complexity is the cumulant crossing method~\cite{Binder+Heermann:1988}.

\begin{figure}[t!]
\centering
\includegraphics[width=8cm]{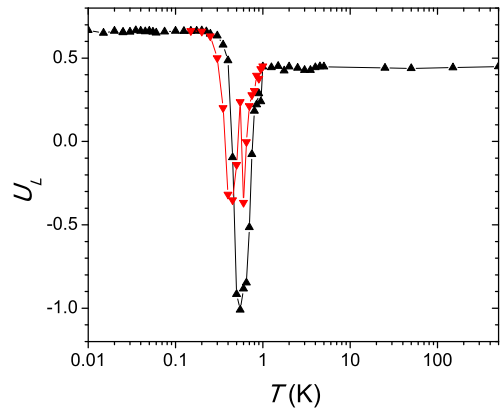}
\caption{An example of the fourth-order cumulant method~\cite{Binder+Heermann:1988} for $x=0.03$, using the 1st set of the $J_{ij}$.
We identify $T_{\mathrm{C}}$ with the intersection of $U_L$ for various linear sizes $L$ .
We show results for two cubes:
$L = $ 50 $a_0$ ($N_{\textrm{Mn}} =$ 15000, down triangles) and
$L = $ 60 $a_0$ ($N_{\textrm{Mn}} =$ 25920, up triangles).
The determined $T_{\mathrm{C}} =  0.5 \pm  0.05$~K.}
\label{Fig:3percent10NNJsTC}
\end{figure}

The {\it fourth-order cumulant} for a lattice of linear size $L$
\begin{equation}\label{ul}
U_L = 1 - \frac{\langle S^4 \rangle_L}{3 \langle S^2 \rangle_L^2}.\vspace{-0.2cm}
\end{equation}
$\langle S^2 \rangle_L$ and $\langle S^4 \rangle_L$ are the statistical averages of the squares and of the fourth powers of $S$, respectively, averages taken over systems at equilibrium at a constant temperature $T$.
$U_L$ has a size-independent catholic fixed point, i.e. all $U_L(T)$ curves for different $L$ cross at $T_{\mathrm{C}}$. In other words, one may plot $U_L(T)$ for various $L$'s and estimate $T_{\mathrm{C}}$
from the common intersection point.
We expect that, according to Eq.~\ref{ul}, for very low temperatures $U_L = \frac{2}{3}$,
and at very high temperatures  $U_L = \frac{4}{9}$.
Although there may be some scatter at the intersection points for different pairs ($L$,$L'$) especially for very small sizes, the accuracy of the fourth-order cumulant method is very satisfactory.
Suppose that we plot $U_L(T)$ and  $U_{L'}(T)$ with $L' > L$.
For $T < T_{\mathrm{C}} \Rightarrow U_{L'} > U_L$  and
for $T > T_{\mathrm{C}} \Rightarrow U_{L'} < U_L$, i.e.
the plots of $U_L(T)$ for a number of sizes should intersect at a point or at least their pairwise intersections should be fairly close: the intersection point is good estimation of $T_{\mathrm{C}}$.
A example of the fourth-order cumulant method~\cite{Binder+Heermann:1988} applied to our system for $x=0.03$ and using the 1st set of $J_{ij}$ is shown in Fig.~\ref{Fig:3percent10NNJsTC}.

%%%%%%%%%%%%%%%%%%%%%%%%%%%%
\section{Results}%%%%%%%%%%%
\begin{figure}[h!]
\centering
\includegraphics[width=8cm]{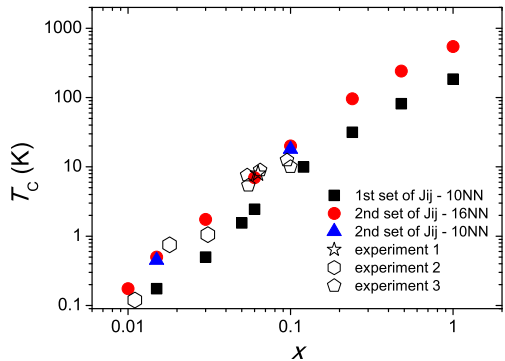}
\caption{Computed values of $T_\mathrm{C}(x)$ (full symbols) compared to experimental values (empty symbols).  Squares, circles, and triangles represent $T_\mathrm{C}(x)$ values obtained by Monte Carlo simulations using the 1st set of $J_{ij}$ (Fig.~\ref{Fig:JsTC}(a)), the 2nd set up to 16th NNs (Fig.~\ref{Fig:JsTC}(b)), and the 2nd set up to 10th NNs, respectively. Empty symbols represent experiments results (star~\cite{sarigiannidou:2006}, hexagons~\cite{sawicki:2012}, and pentagons~\cite{stefanowicz:2013}).}
\label{Fig:JsTC_exp}
\end{figure}

\label{RD}%%%%%%%%%%%%%%%%%%
%%%%%%%%%%%%%%%%%%%%%%%%%%%%
In Fig.~\ref{Fig:JsTC_exp} we show magnitudes of $T_\mathrm{C}$ obtained by MC simulations using the sets of $J_{ij}$ values shown in Figs.~\ref{Fig:JsTC}(a) and \ref{Fig:JsTC}(b). The computation results for the 2nd set are in remarkable agreement with the available experimental data~\cite{sawicki:2012,stefanowicz:2013,sarigiannidou:2006}.
The 2nd set of $J_{ij}$ values, obtained for a smaller value of $e_2$, leads to improve agreement with the experimental data, and allows to compute $T_\mathrm{C}(x)$ down to $x = 0.01$. As shown, the use of 10 NNs within the 2nd set (up triangles) results already in accurate magnitudes of $T_\mathrm{C}$. Assuming no insulator-to-metal transition, i.e. the absence of delocalized holes, we predict room-temperature ferromagnetism for $ x \gtrsim  0.5$ in Ga$_{1-x}$Mn$_x$N with randomly distributed Mn$^{3+}$ ions.

%%%%%%%%%%%%%%%%%%%%%%%%%
\section{Conclusion}%%%%%
\label{conclusion}%%%%%%%
%%%%%%%%%%%%%%%%%%%%%%%%%
Our theoretical results on $T_\mathrm{C}(x)$ in Ga$_{1-x}$Mn$_x$N agree quantitatively with the measured values in the experimentally explored range $ 0.01 \leq x \leq 0.1$~\cite{sawicki:2012,stefanowicz:2013,sarigiannidou:2006}. This agreement supports the view that ferromagnetic superexchange is the dominant coupling mechanism between Ga-substitutional Mn$^{3+}$ ions in Ga$_{1-x}$Mn$_x$N, leading to $T_\mathrm{C} \approx$ 13~K at $x = 0.1$. The ferromagnetic character of the coupling is in accord with the Anderson-Goodenough-Kanamori rules for partly filled $t_2$ states of tetrahedrally coordinated TM ions. Furthermore, the identical value of the exponent $m = 2.2\pm 0.2$ in the dependence $T_\mathrm{c}(x) \propto x^m$ both for ferromagnetic ordering as found in Ga$_{1-x}$Mn$_x$N, and spin-glass freezing observed in Mn- and Co-based II-VI DMSs (in which $t_2$ states are entirely occupied for the majority spin direction)~\cite{spin-glass-freezing,Sawicki:2013_PRB} verifies the scaling law \cite{rammal:1982} implying that independently of the sign of $J_{ij}$, $m = \lambda/d$, where $J_{ij} \propto R_{ij}^{-\lambda}$ and $d$ is the space dimensionality.

According to our theoretical model for randomly distributed Mn$^{3+}$ ions over cation sites,
room-temperature ferromagnetism will appear for $x \gtrsim 0.5$, if the high-$T_\mathrm{C}$ regime will not be shifted to even lower $x$ values by the insulator-to-metal transition and the associated delocalization of holes supplied by Mn ions~\cite{Dietl:2008_PRB}. A future growth effort will show whether it is possible to obtain Ga$_{1-x}$Mn$_x$N with
merely randomly distributed Mn$^{3+}$ in a concentration $x$ sizably exceeding 0.1.

Comparing to the {\it ab initio} data \cite{sato:2010}, our $J_{ij}$ magnitudes are significantly smaller and result in a stronger dependence of $T_\mathrm{C}$ on Mn content $x$. This suggests that the current first principles methods overestimate coupling between TM levels and band states, which--in turn--is adequately taken into account within the Parmenter's generalization of the Anderson Hamiltonian \cite{blinowski:1996} employed here. Furthermore, since within the present formalism it is possible to compute the magnitudes of $J_{ij}$ for virtually any distance $R_{ij}$, we have been able to evaluate $T_\mathrm{C}(x)$ down to the experimentally relevant range of $x = 0.01$.

\begin{acknowledgments}
This work was supported by the FunDMS Advanced Grant of the ERC (Grant No. 227690) within the Ideas 7th Framework Programme of European Community.
Computer time in Athens was partly provided by the National Grid Infrastructure HellasGrid.
\end{acknowledgments}

\end{document}